\documentclass[twocolumn,showpacs,amsmath,amssymb,pre]{revtex4-1}

\usepackage{graphicx}
\usepackage{color}
\usepackage{dcolumn}
\usepackage{amsmath}
\usepackage{subfigure}

\usepackage{sidecap}
\usepackage{dsfont}

\usepackage{mathrsfs}
\usepackage{amsfonts}
\usepackage{indentfirst}
\usepackage{bm}
\usepackage{bbm}

\usepackage[colorlinks,citecolor=blue]{hyperref}

\newcommand{\be}{\begin{equation}}
\newcommand{\ee}{\end{equation}}
\newcommand{\bey}{\begin{eqnarray}}
\newcommand{\eey}{\end{eqnarray}}
\newcommand{\bw}{\begin{widetext}}
\newcommand{\ew}{\end{widetext}}

\newcommand{\ra}{\rangle}
\newcommand{\la}{\langle}

\newcommand{\ba}{\begin{array}}
\newcommand{\ea}{\end{array}}
\newcommand{\bi}{\begin{itemize}}
\newcommand{\ei}{\end{itemize}}
\newcommand{\bem}{\begin{enumerate}}
\newcommand{\eem}{\end{enumerate}}

\begin{document}

\title{Power-law decay of the fraction of the mixed eigenstates in kicked top model 
with mixed-type classical phase space}

\author{Qian Wang}
\affiliation{CAMTP-Center for Applied Mathematics and Theoretical Physics, University of Maribor, 
Mladinska 3, SI-2000 Maribor, Slovenia, European Union, and \\
Department of Physics, Zhejiang Normal University, Jinhua 321004, China}

\author{Marko Robnik}
\affiliation{CAMTP-Center for Applied Mathematics and Theoretical Physics, University of Maribor, 
Mladinska 3, SI-2000 Maribor, Slovenia, European Union}

\begin{abstract}

The properties of mixed eigenstates in a generic quantum system with classical counterpart 
that has mixed-type phase space, although important to understand several 
fundamental questions that arise in both theoretical and experimental studies, are still not clear.
Here, following a recent work [\v{C}.~Lozej {\it et al}. Phys. Rev. E {\bf 106}, 054203 (2022)], we perform 
an analysis of the features of mixed eigenstates in a time-dependent Hamiltonian system, 
the celebrated kicked top model. 
As a paradigmatic model for studying quantum chaos, kicked top model is known to 
exhibit both classical and quantum chaos.  
The types of eigenstates are identified by means of the phase space overlap index, 
which is defined as the overlap of the Husimi function with
regular and chaotic regions in classical phase space.
We show that the mixed eigenstates appear due to various tunneling precesses 
between different phase space structures, while the regular and chaotic eigenstates are, respectively, 
associated with invariant tori and chaotic component in phase space. 
We examine how the probability distribution of the phase space overlap index evolves with increasing
system size for different kicking strengths.
In particular, we find that the relative fraction of mixed states exhibits a power-law decay as the system size increases,
indicating that only purely regular and chaotic eigenstates are left in the strict semiclassical limit.
We thus provide further verification of the principle of uniform semiclassical condensation of Husimi functions and
confirm the correctness of the Berry-Robnik picture.

\end{abstract}

\date{\today}

\maketitle

\section{Introduction}

The pivotal role played by the quantum chaos in studying various important 
questions in numerous branches of physics has
triggered a great deal of efforts to explore different aspects of quantum chaos 
\cite{Izrailev1990,Stockmann1999,Haake2001,Ullmo2008,Gomez2011,Borgonovi2016,Alessio2016,
Jahnke2019,Lucas2020,Dymarsky2020}.
However,  a full understanding of the properties of the quantum systems 
associated with classical mixed-type systems is still lacking.
Classically, the mixed-type systems exhibit both regular and chaotic motion and result in an 
intricate hierarchical structure in their phase space, with regular islands embedded 
in the chaotic sea \cite{Lichtenberg2013}. 
This lead Percival to conjecture that the eigenstates in the corresponding quantum systems 
should be of either the regular or chaotic type \cite{Percival1973}. 
With further elaboration made by Berry \cite{Berry1977a,Berry1977b}, this conjecture finally develops into
the so-called principle of uniform semiclassical condensation 
of Wigner functions (or Husimi functions) (PUSC) \cite{Robnik1998}. 
For details see recent review papers \cite{Robnik2019,Robnik2020} and references therein.

According to the PUSC, the eigenstates in a generic quantum system are either  
condensed on the invariant tori in the regular islands, referred as the regular states, 
or supported on the chaotic sea, known as the chaotic states, 
in the ultimate semiclassical limit where the classical action is much larger than the Planck constant.
Consequently, the spectral statistics for the regular and chaotic states are separately described by the Poissonian
statistics \cite{Berry1977} and random matrix theory (RMT) \cite{BGS1984,Wigner1993,Mehta2004}, 
while the whole spectrum is well captured by the Berry-Robnik (BR) picture \cite{Berry1984}.
The validity of the BR distribution to characterize the spectral statistics in generic quantum systems has been
numerically verified by numerous works 
\cite{Prosen1993b,Prosen1993b,Prosen1994a,Prosen1994b,Baowen1994,Prosen1995,Baowen1995,
Prosen1998,Prosen1999,Veble1999,Manos2013}.
However, in the near semiclassical limit, it is natural to expect that there should 
be an intermediate regime in which many eigenstates will behave as mixed states 
due to various tunneling processes between different phase space structures.
Although the mixed eigenstates exhibit several important and interesting phenomena, 
such as chaos-assisted tunneling \cite{Tomsovic1994} which has potential applications 
in quantum simulation \cite{Martinez2021} and can be used to create highly entangled states \cite{Vanhaele2022}, 
much of their properties remain unknown. 
  
Very recently, using the Husimi function \cite{Husimi1940}, 
the properties of the mixed states 
in the lemon billiards \cite{Heller1993,Makino2001,Lozej2021a,Lozej2021b,Lozej2021c} 
have been explored \cite{Lozej2022}.
In the present work, we continue and extend this study to provide a detailed investigation 
of the signatures of the mixed states in the kicked top model 
\cite{Haake1987}, a paradigmatic model in the studies of quantum chaos \cite{Haake2001}, which
has been realized in a variety of experimental platforms, 
such as cold atoms \cite{Chaudhury2009}, superconducting circuits \cite{Neill2016}, 
and nuclear magnetic resonance simulator \cite{Krithika2019}. 
As the kicked top model exhibits a transition from the regular regime to 
the chaotic one with increasing kicking strength \cite{Haake2001,Haake1987}, 
it therefore provides us a model system to analyze the features of the mixed states.
 
Following the method used in Ref.~\cite{Lozej2022}, the mixed states are identified by 
the phase space overlap index defined through the Husimi functions of the eigenstates.
We show that the probability distribution of the phase space overlap index has double-peak shape and
bears a remarkable change as the semiclassical limit is approached (namely, by increasing the system size).
More insights about the properties of the mixed states are gained from dependence of their 
proportion on the system size.
We demonstrate that the proportion of mixed states which belong to a certain interval of the phase space 
overlap index, follows a power-law decay with increasing system size.
This confirms the disappearance of the relative fraction of the mixed states in the semiclassical limit, 
as unveiled in lemon billiards and in consistent with the PUSC. 
It further verifies the correctness of the Berry-Robnik picture for describing 
the spectral statistics in generic quantum systems.

The structure of the article is the following.
In Sec.~\ref{Second}, we introduce the kicked top model and 
analyze the integrability-chaos transition for classical
and quantum cases by means of the largest Lyapunov exponent and 
Kolmogorov-Sinai entropy, as well as the spectral statistics, respectively. 
In addition, the definition and calculation of the Husimi function for an individual eigenstate 
are also discussed in this section. 
Then, in Sec.~\ref{Third}, we perform an explicit analysis of the 
probability distribution of the phase space overlap index, showing
how it evolves with increasing system size.
In this section, we further examine the dependence of the proportion of mixed states on the system size.
We finally conclude in Sec.~\ref{Fourth} with several remarks.

 \begin{figure*}
  \includegraphics[width=\textwidth]{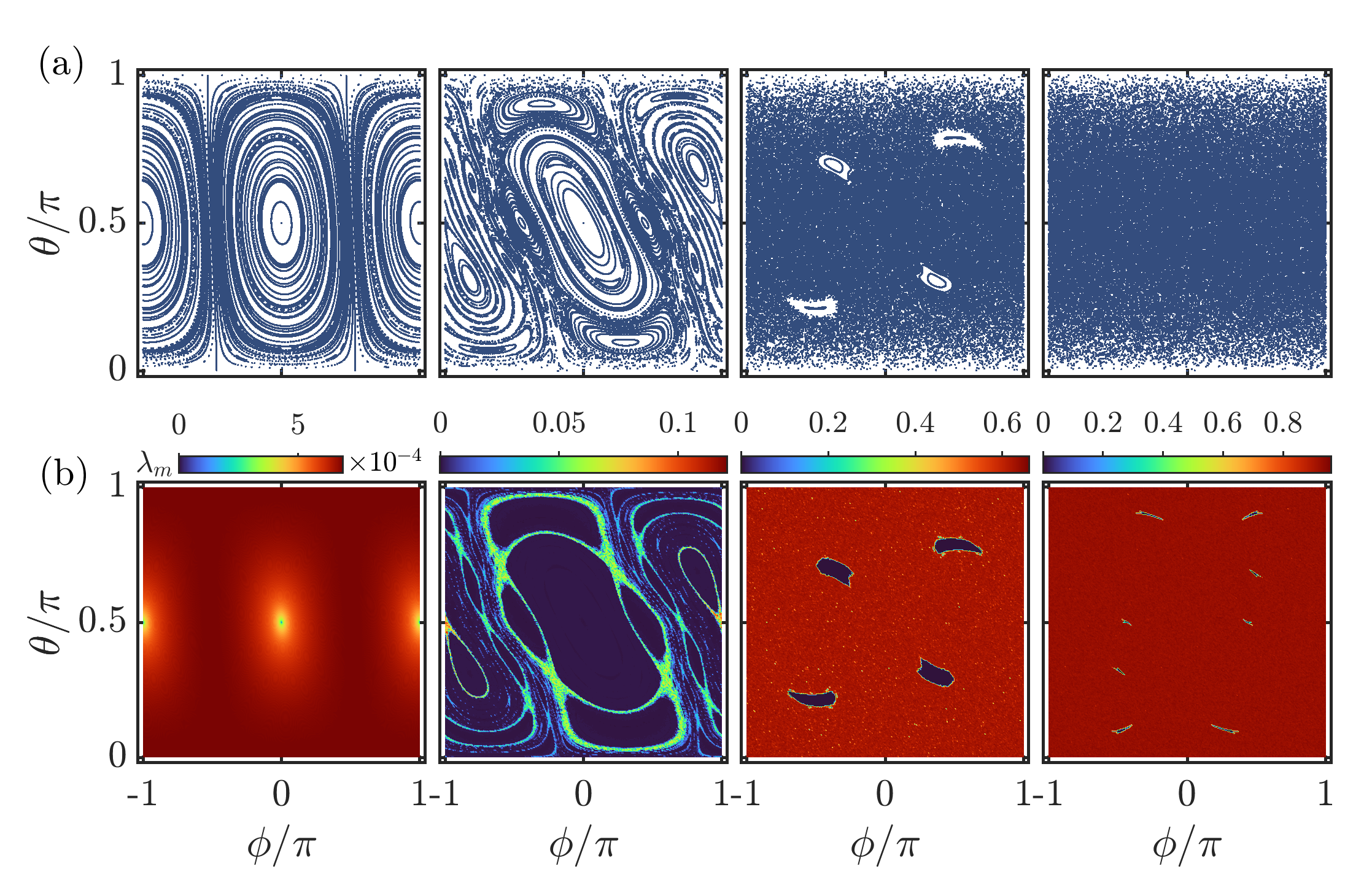}
  \caption{(a) Classical phase space portraits with $225$ random initial conditions for (from left to right) 
  $\gamma=0.2, 2, 4$, and $\gamma=6$.
  Each initial condition has been evolved for $400$ kicks. 
  (b) Largest Lyapunov exponent, $\lambda_m$, of the classical kicked top model for the same values of $\gamma$ as in (a).
  The largest Lyapunov exponent has been calculated on a grid of $300\times300$ initial conditions, 
  each has a duration of $1\times10^4$ kicks. 
  Other parameter: $\alpha=11\pi/19$.}
  \label{PSKS}
 \end{figure*}

\section{Kicked top model} \label{Second}

The model we study is the kicked top model, which is a time dependent system with the Hamiltonian
given by (setting $\hbar=1$) \cite{Haake1987}
\be \label{KTH}
   H=\alpha J_x+\frac{\gamma}{2j}J_z^2\sum_{n=-\infty}^{+\infty}\delta(t-n),
\ee 
where $J_\nu$ with $\nu=x,y,z$ are the angular momentum operators of the total spin $j$ system.
$\alpha$ denotes the angle of the precession around $x$-axis, and $\gamma$ 
is the strength of the kicking with a period that we have set to unity. 
It is worth pointing out that the dependence of both quantum and classical dynamics 
of the model on the value of $\alpha$ has been investigated in our previous work \cite{WangQ2021}.
Here, we have checked that our main results are independent of the specific value of $\alpha$.
We thus fixed $\alpha=11\pi/19$ throughout this work.

The dynamical evolution under above Hamiltonian is governed by the Floquet operator
\be \label{FOP}
  F=e^{-i\frac{\gamma}{2j}J_z^2}e^{-i\alpha J_x}.
\ee
One can easily find that the Hamiltonian (\ref{KTH}) conserves the total spin $j$. 
Hence, the Hilbert space of the system has dimension $\mathcal{D}_\mathcal{H}=2j+1$.
In our numerical calculation, the basis for the Hilbert space is the Dicke states, $\{|j,m\ra\}_{m=-j}^{m=+j}$, satisfying
$J_z|j,m\ra=m|j,m\ra$ and $\mathbf{J}^2|j,m\ra=j(j+1)|j,m\ra$ with $\mathbf{J}^2=J_x^2+J_y^2+J_z^2$. 
Then, the elements of the Floquet operator are 
\be
  \la j,m|F|j,m'\ra=\exp\left({-i\frac{\gamma}{2j}m^{ 2}}\right)\mathcal{W}_{mm'},
\ee
where $\mathcal{W}_{mm'}$ is the Winger $D$ function \cite{Rose1995} and can be calculated as
\begin{align}
  \mathcal{W}_{mm'}&=\la j,m|e^{-i\alpha J_x}|j,m'\ra \notag \\
       &=\sum_{k_x=-j}^{k_x=+j}e^{-i\alpha k_x}\la j,m|j,k_x\ra\la j,k_x|j,m'\ra,
\end{align}
with $|j,k_x\ra$ representing the eigenstates of $J_x$, so that, $J_x|j,k_x\ra=k_x|j,k_x\ra$. 

The time evolution of the angular momentum is given by the map
$J_\nu(n+1)=F^\dag J_\nu(n)F$, which can be explicitly written as \cite{Haake1987,Fox1994,Arias2021}
\begin{align} \label{Hmap}
\begin{aligned}
  &J_x(n+1)=\frac{1}{2}\left\{J_x(n)+i\Theta_n(\alpha)\right\}\exp\left[i\frac{\gamma}{2j}\Xi_n(\alpha)\right]+\mathrm{h.c.}, \\
  &J_y(n+1)=\frac{1}{2i}\left\{J_x(n)+i\Theta_n(\alpha)\right\}\exp\left[i\frac{\gamma}{2j}\Xi_n(\alpha)\right]+\mathrm{h.c.}, \\
  &J_z(n+1)=J_y(n)\sin\alpha+J_z(n)\cos\alpha,
\end{aligned}
\end{align} 
where $\Theta_n(\alpha)=J_y(n)\cos\alpha-J_z(n)\sin\alpha$ and 
$\Xi_n(\alpha)=2[J_y(n)\sin\alpha+J_z(n)\cos\alpha]+1$.
A detailed derivation of above equation can also be found in appendix of Ref.~\cite{WQian2023}.
  
\subsection{Classical kicked top model}

The classical counterpart of the kicked top model is obtained by taking the classical limit $j\to\infty$, 
which means that one can define an effective Planck constant as $\hbar_{\mathrm{eff}}=1/j$. 
To obtain the classical equations of motion of the kicked top model, we first introduce the normalized vector
$\mathbf{X}=\la\mathbf{J}\ra/j$, which becomes a classical vector when $j\to\infty$.
Then, as the expectation value of the products of the evolved angular momentum 
operators in Eq.~(\ref{Hmap}) can be factorized as $\la J_\mu J_\nu\ra=\la J_\mu\ra\la J_\nu\ra$ in the classical limit, 
it is straightforward to show that the classical map for the classical vector $\mathbf{X}$ takes the form 
\cite{Arias2021,Piga2019}
\begin{align} \label{Clasmap}
\begin{bmatrix}
X_{n+1} \\
Y_{n+1} \\
Z_{n+1}
\end{bmatrix}  
=
\begin{bmatrix}
\cos\Omega_n & -\cos\alpha\sin\Omega_n & \sin\alpha\sin\Omega_n \\
\sin\Omega_n & \cos\alpha\cos\Omega_n & -\sin\alpha\cos\Omega_n \\
0 & \sin\alpha & \cos\alpha
\end{bmatrix} 
\begin{bmatrix}
X_n \\
Y_n \\
Z_n
\end{bmatrix},
\end{align}
where $\Omega_n=\gamma(Y_n\sin\alpha+Z_n\cos\alpha)$.
The normalization of $\mathbf{X}$ allows us to parametrize it as 
$\mathbf{X}=(\cos\phi\sin\theta, \sin\phi\sin\theta,\cos\theta)$ with $\theta$ and $\phi$ being 
the azimuthal and polar angles, respectively.
Hence, the classical phase space can be described by variables $\phi=\arctan(Y/X)$ and $\theta=\arccos(Z)$. 

 \begin{figure}
  \includegraphics[width=\columnwidth]{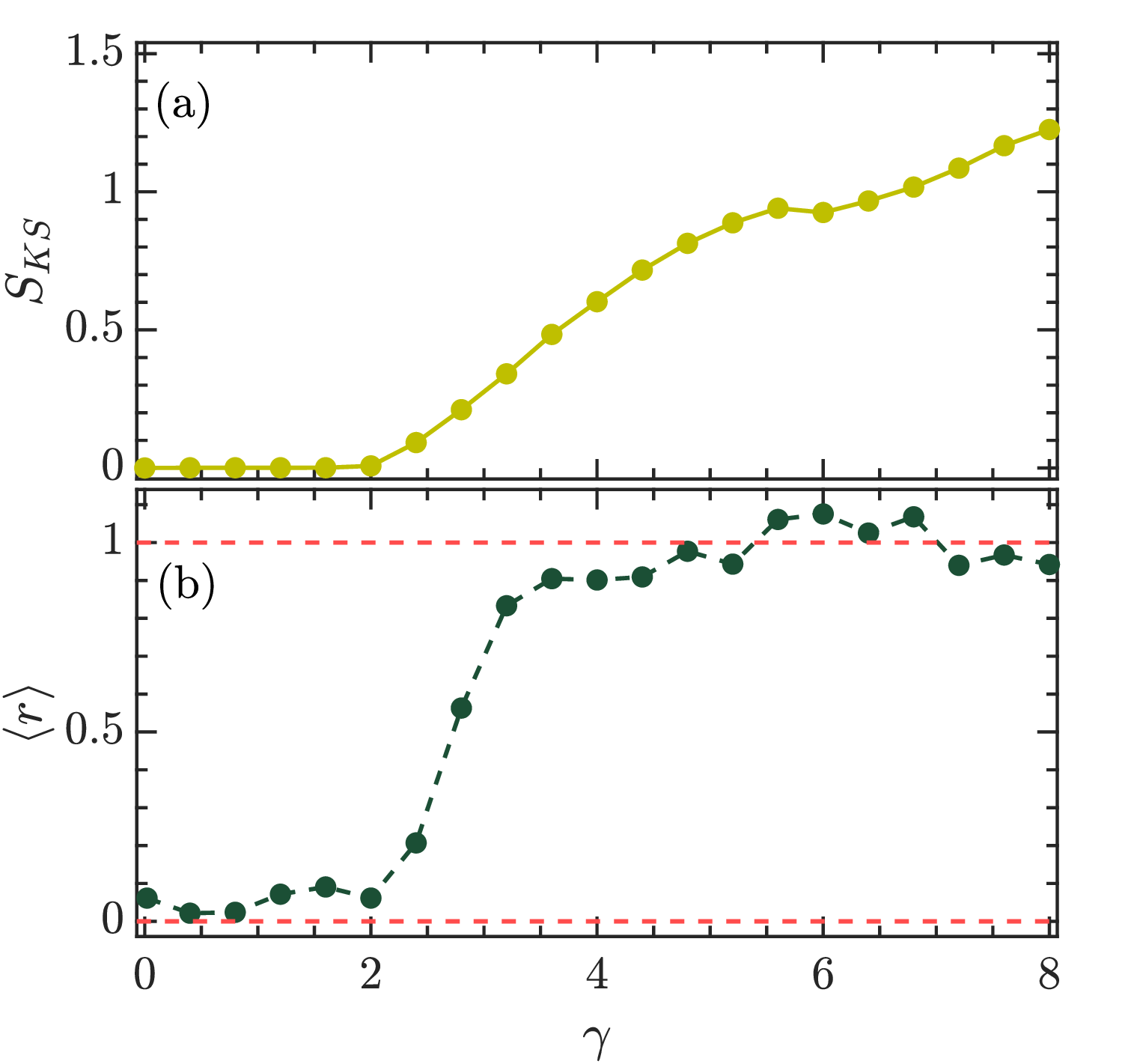}
  \caption{(a) Kolmogorov-Sinai (KS) entropy, ${S}_{KS}$, as a function of kicking strength $\gamma$.
  ${S}_{KS}$ is obtained by averaging over $90000$ initial points in the phase space, 
  each evolved for $1\times10^4$ kicks. 
  (b) Variation of rescaled average level spacing ratio with $\gamma$ for $j=2500$.
  The upper and bottom red dashed lines denote $\widetilde{\la r\ra}=1$ and $0$, respectively.
  Other parameter: $\alpha=11\pi/19$.}
  \label{KSRavg}
 \end{figure}

It is known that the classical map in Eq.~(\ref{Clasmap}) undergoes a transition 
from integrability to chaos with increasing kicking strength $\gamma$. 
This is demonstrated in Fig.~\ref{PSKS}(a), where we plot the Poincar{\'e} section of the classical top model 
for different $\gamma$ values.
It can be clearly seen that the classical phase space is dominated by regular orbits for small $\gamma$ 
and turns into the mixed dynamics with regular islands embedded in the chaotic sea as $\gamma$ is increased.  
The regular islands disappear for even larger $\gamma$ and the phase space is fully covered by the chaotic sea,
as shown in the right most column of Fig.~\ref{PSKS}(a).  

To quantitatively capture the chaotic transition illustrated in Fig.~\ref{PSKS}(a), we consider the
largest Lyapunov exponent, which decribes the rate of the deviation between two 
initially nearby close orbits and can be calculated as \cite{Lichtenberg2013,Piga2019}
\be
   \lambda_m=\lim_{t\to\infty}\frac{1}{t}\sum_{n=1}^t\ln d_n,
\ee 
where $d_n=[(\delta X_n)^2+(\delta Y_n)^2+(\delta Z_n)^2]^{1/2}$ is the phase space distance between two initially 
nearby points after $n$ kicks.
Here, $\delta\mathbf{X}$ is determined by the tangent map \cite{Piga2019},
$\delta\mathbf{X}_{n+1}=\left[\partial\mathbf{X}_{n+1}/\partial\mathbf{X}_n\right]\delta\mathbf{X}_n$,
with initial condition $\delta\mathbf{X}_0$.
Moreover, we renormalize $d_n$ at each step $n$ in our calculation. 
As the largest Lyapunov exponent quantifies how two infinitesimally orbits separate with time, 
it therefore acts as a measure of the level of chaos. 
For the regular regions in the phase space, we have $\lambda_m=0$, 
while $\lambda_m>0$ for the chaotic component.

 \begin{figure*}
  \includegraphics[width=\textwidth]{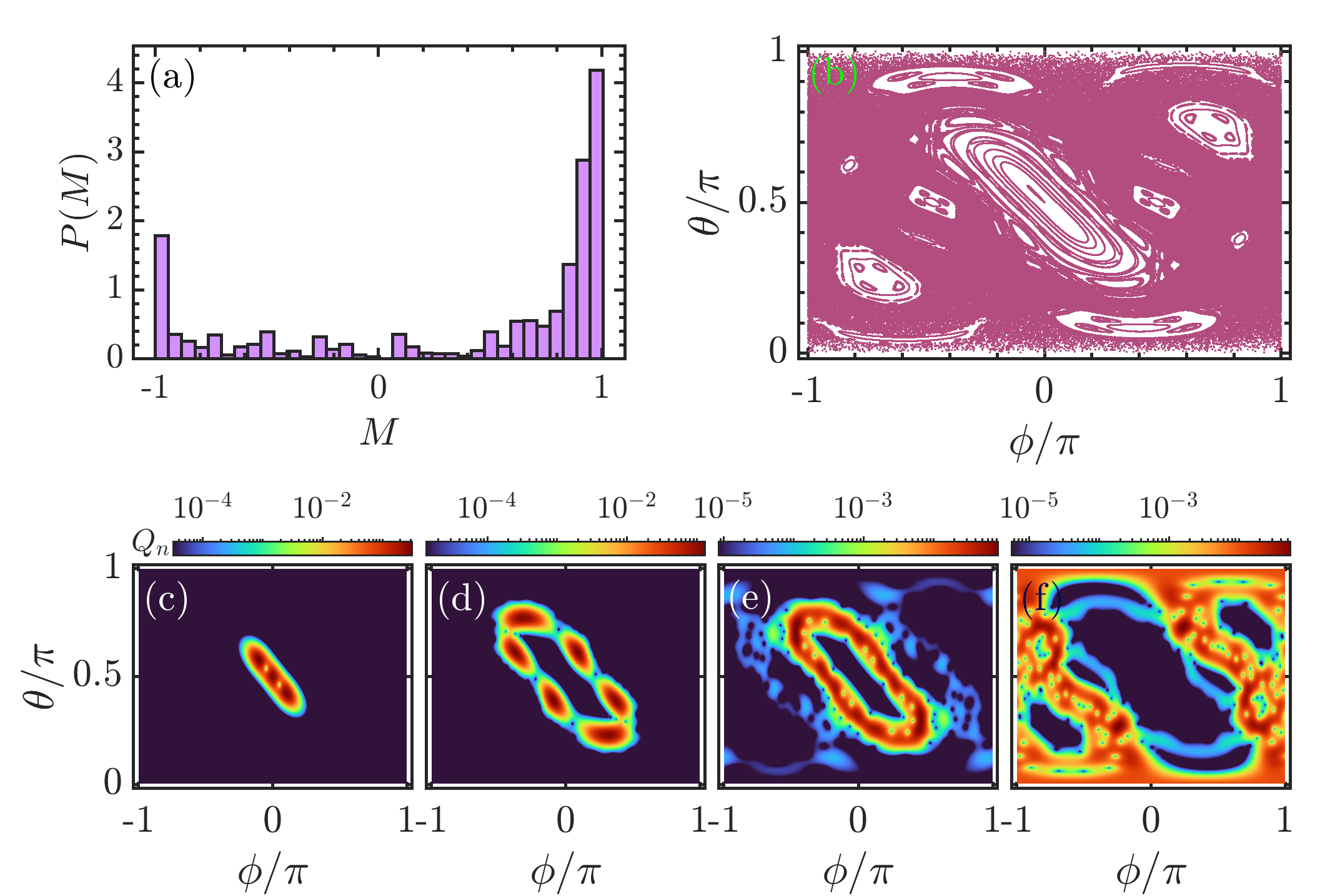}
  \caption{(a) Histogram of the probability distribution $P(M)$ for the kicked top model with $\gamma=2.6$.
  The histogram is obtained for the eigenstates of the system with sizes $j\in[150,154]$, increasing in step of $1$. 
  (b) Poincar\'{e} section of the kicked top model for $\gamma=2.6$.
  (c)-(f): Husimi functions for the eigenstates $n=5, M_n=-1$ (c), $n=36, M_n=-0.5161$ (d), $n=10, M_n=0.3075$ (e), and
  $n=264, M_n=0.9744$ (f) with $j=150$ and $\gamma=2.6$. 
  Other parameter: $\alpha=11\pi/19$.}
  \label{MDis}
 \end{figure*}

Figure \ref{PSKS}(b) plots the largest Lyapunov exponent for the same values of $\gamma$ as in Fig.~\ref{PSKS}(a).   
A remarkable resemblance between Fig.~\ref{PSKS}(a) and \ref{PSKS}(b) can be obviously observed.  
Specifically, the very tiny $\lambda_m$ at small $\gamma$ is in agreement 
with the regular dynamics, as seen in the first column of Fig.~\ref{PSKS}.  
For the mixed phase space, one can see that the $\lambda_m=0$ regions are clearly corresponding to the regular regions 
in the phase space, while the chaotic regions are marked by $\lambda_m>0$, 
as demonstrated in the second and third columns of Fig.~\ref{PSKS}.
The globally chaotic dynamics at $\gamma=6$ results in the larger values 
and an almost uniform distribution of the largest Lyapunov exponent 
in the phase space (see the last column of Fig.~\ref{PSKS}).
In particular, the largest Lyapunov exponent can help us identify 
the invisible regular islands in the Poincar{\'e} section.  

Further characterizations of the integrability-chaos transition in the classical kicked top model are revealed 
by the Kolmogorov-Sinai (KS) entropy, which is, generally speaking, 
related to the rate of change of the coarse-grained 
Gibbs entropy with time \cite{Kolmogorov1959} and for the kicked top model is calculated as 
\cite{Lichtenberg2013,Pesin1977} 
\be
   S_{KS}=\frac{1}{4\pi}\int\lambda_md\mathcal{A},
\ee
where $d\mathcal{A}=\sin\theta d\theta d\phi$ is the phase space area element \cite{Ariano1992}. 
Figure \ref{KSRavg}(a) shows how the KS entropy, ${S}_{KS}$, varies with
increasing $\gamma$. 
The value of ${S}_{KS}$ remains zero for $\gamma\leq2$ and begins to grow at $\gamma>2$.
This means that the model undergoes a chaotic transition around $\gamma=2$ 
and the degree of chaoticity is enhanced with increasing $\gamma$, in accord with 
the phase space features shown in Fig.~\ref{PSKS}(a). 
Here, we would like to point out that the plateau in the behavior of ${S}_{KS}$ around $\gamma\approx6$ 
is due to the existence of tiny regular regions in the classical phase space [see the last figure in Fig.~\ref{PSKS}(b)].

\subsection{Quantum chaos in kicked top model}

The onset of chaos in the classical kicked top model gets reflected 
in its quantum counterpart, resulting in the quantum chaos. 
There are many different ways to diagnose the presence of quantum chaos 
\cite{Guhr1998,Emerson2002,Hurtubise2020,Zonnios2022,Wimberger2022,Garcia2023,Hashimoto2023}.
Among them, the statistics of the spacings $s$ between consecutive energy levels is the most commonly used probe.  
The distribution of $s$ in the chaotic systems is well described by the Wigner surmise \cite{Wigner1993,BGS1984}, 
whereas the regular systems are generically characterized by the Poisson distribution 
\cite{Berry1977,RobnikV1998}. 
Here, instead of analyzing the level spacing distribution, we focus on the 
spacing ratio for three successive levels, first introduced in Ref.~\cite{Oganesyan2007}.  
The big advantage to consider the level spacing ratios rather than level spacings themselves is that 
it avoids the intricate unfolding procedure.  
As a consequence, it becomes the most popular chaos indicator in various studies
\cite{Atas2013,Alessio2014,Giraud2022,Garcia2018,Sierant2019,Corps2020,Lucas2020,
Moudgalya2021,WQian2022,WQian2023,Mateos2023}, 
in particular for many body quantum systems.

For the time-dependent Hamiltonian, such as our studied model, the level spacing ratios are defined as
\be
   r_n=\mathrm{min}\left(\frac{s_n}{s_{n-1}},\frac{s_{n-1}}{s_n}\right),
\ee
where $s_n=\nu_n-\nu_{n-1}$, with $\nu_n$ being the $n$th quasienergy (eigenphase) 
of the Floquet operator in Eq.~(\ref{FOP}).
Clearly, $r$ is defined in the range $0\leq r\leq 1$. 
The distribution $P(r)$ of $r$ for both integrable and chaotic systems have been analytically derived 
\cite{Atas2013,AtasE2013}, from which one can find that the mean level spacing ratio, $\la r\ra=\int_0^1rP(r)dr$, 
behaves as an efficient detector of quantum chaos. 
For integrable systems, one has $\la r\ra_{RG}=2\ln2-1\approx0.386$ \cite{Atas2013}, while 
$\la r\ra_{COE}\approx0.527$ \cite{Alessio2014,Giraud2022} for the fully chaotic systems belonging to 
the circular orthogonal ensemble (COE), such as the kicked top model at $\gamma\simeq6$.

A more convenient quantity that is used to detect the crossover from integrability to quantum chaos is the rescaled mean
level spacing ratio \cite{WQian2023,Lydzba2022}, defined as
\be
  \widetilde{\la r\ra}=\frac{|\la r\ra-\la r\ra_{RG}|}{\la r\ra_{COE}-\la r\ra_{RG}}.
\ee
It varies in the interval $\widetilde{\la r\ra}\in[0,1]$. 
When $\widetilde{\la r\ra}=0$, it indicates the regular dynamics in the system.
On the contrary, the fully chaotic dynamics in the system leads to $\widetilde{\la r\ra}=1$.   
In Fig.~\ref{KSRavg}(b), we display how $\widetilde{\la r\ra}$ evolves as a function of $\gamma$. 
We can see that the transition of $\widetilde{\la r\ra}$ from a value close to zero 
to a value around one confirms the onset of chaos as $\gamma$ is increased. 
Moreover, the agreement between the onset of chaos in ${S}_{KS}$ 
and $\widetilde{\la r\ra}$ at $\gamma\approx2$
further corroborates a good quantum-classical correspondence.

\subsection{Husimi function} 

Our aim is to explore the properties of the eigenstates in a 
quantum system with mixed phase space in the classical limit.
It is therefore required to identify the various types of the eigenstates.
As in previous works \cite{Batistic2013a,Batistic2013b,Robnik2016,Lozej2018,Robnik2020}, 
we use the Husimi function \cite{Husimi1940} to characterize 
the signatures of the eigenstates in classical phase space.

The Husimi function can unveil various aspects of the eigenstates exhibited in the phase space, 
in particular their localization properties
\cite{WQian2023,QWang2020,QWang2021,Villasenor2021,Pilatowsky2022a,Pilatowsky2022b,Lozej2022}.  
To define the Husimi function for the kicked top model, we first introduce the 
generalized $\mathrm{SU}(2)$ spin coherent states, defined as
a rotation of the Dicke state $|j,j\ra$, which can be explicitly written as 
\cite{Radcliffe1971,Perelomov1977,ZhangWM1990}
\begin{align}
  |\phi,\theta\ra&=e^{i\theta(J_x\sin\phi-J_y\cos\phi)}|j,j\ra \notag \\   
        &=\sum_{m=-j}^{+j}\frac{\xi^{j-m}}{(1+|\xi|^2)^j}\sqrt{\frac{(2j)!}{(j+m)!(j-m)!}}|j,m\ra,
\end{align}
where $\xi=\tan(\theta/2)e^{i\phi}$ with $\phi\in[-\pi,\pi)$ and $\theta\in[0,\pi]$. 
The over completeness of the coherent states results in the following closure relation
\be
   \mathbb{I}=\frac{2j+1}{4\pi}\int|\phi,\theta\ra\la\phi,\theta|\sin\theta d\theta d\phi.
\ee  
Then the Husimi function for the $n$th eignestate, $|\nu_n\ra$, of $F$ in (\ref{FOP}) is given by
\be
    Q_n(\phi,\theta)=|\la\theta,\phi|\nu_n\ra|^2,
\ee
with the normalization condition 
\be
   \frac{2j+1}{4\pi}\int Q_n(\phi,\theta)\sin\theta d\theta d\phi=1.
 \ee
 
 The principle of uniform semiclassical condensation of the Wigner and Husimi functions predicts
 that the Husimi functions will condense either on the classically invariant torus or chaotic regions in
 the semiclassical limit. 
 However, before the semiclassical limit is reached in practice, one expects that there should 
 exist mixed eigenstates with associated Husimi functions occupying both regular and chaotic regions. 
 In the following section, we identify these mixed states by means of the phase space overlap index and
 discuss how their relative fraction varies as the semiclassical limit is approached. 
 We shall demonstrate that it decays as a power law.

  \begin{figure*}
  \includegraphics[width=\textwidth]{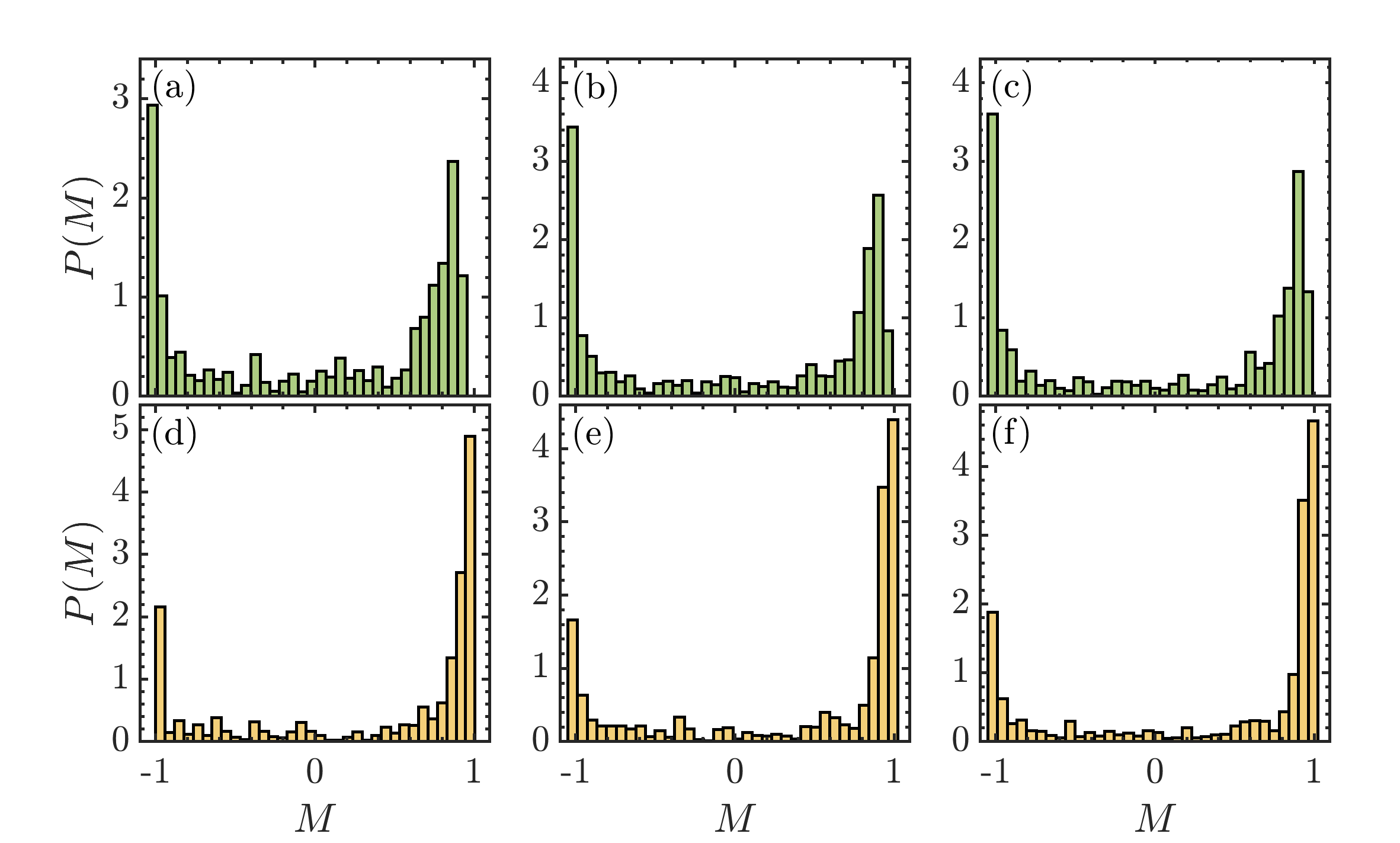}
  \caption{(a)-(c): Histograms of $P(M)$ for different ensembles of the system size: (a) $j\in[200,204]$, (b) $j\in[300,304]$,
  and (c) $j\in[400,404]$. The kicking strength is $\gamma=2.3$.
  (d)-(f): Histograms of $P(M)$ for the same system size ensembles as in panels (a)-(c) with $\gamma=2.6$. 
  Other parameter: $\alpha=11\pi/19$. The system size $j$ in each ensemble is increased in step of $1$.}
  \label{PMdNk}
 \end{figure*}

\section{Phase space overlap index} \label{Third}

Armed with the Husimi function, let us discuss how to identify the types of the eigenstates using approach
that has been performed in Refs.~\cite{Robnik2016,Batistic2013a,Lozej2022}.
We divide the classical phase space $(\phi,\theta)$ into a grid with cells of equal area. 
Each cell is marked by its center point with index $(i,j)$. 
We then define a discrete quantity $C_{ij}$, which takes value $+1$ if the grid point $(i,j)$ 
resides in the chaotic regions and $-1$ otherwise.
Accordingly, the Husimi function of the $n$th eigenstate is discretized on the grid and 
normalized as $[(2j+1)\pi/(2N)]\sum_{i,j}\sin\theta_jQ_n(\phi_i,\theta_j)=1$ 
with $N$ being the number of grid points.

To elucidate whether the $n$th eigenstate is the regular or chaotic eigenstate, we define an overlap index
\be
    M_n=\frac{(2j+1)\pi}{2N}\sum_{i,j}\sin\theta_j Q_n(\phi_i,\theta_j)C_{ij}.
\ee
In the ultimate semiclassical limit, one can expect that $M$ should take the value either $-1$ or $+1$, 
corresponding to the regular or chaotic eigenstates, respectively. 
However, since the semiclassical limit is not yet reached in practice, 
$M$ actually varies between $-1$ and $+1$.
Hence, it is natural to ask what is the distribution of $M$ and 
how it changes as the semiclassical limit is approached.

Previously, the joint distribution of $M$ and phase space localization measures 
has been analyzed in the billiard systems and it was found that the distribution of $M$ 
turns into a double peak distribution when approaching the semiclassical limit \cite{Lozej2022}. 
In addition, approaching the semiclassical limit also led to a power-law decay of 
the proportion of the mixed states with intermediate values of $M$. 
In the following of this section, we address above mentioned questions 
in the kicked top model and provide further
evidence of the power-law decay exhibited by the 
fraction of the mixed states in the semiclassical limit.

Let us first consider the probability distribution of $M$, which is defined as
\be
   P(M\in\Lambda_n)=\frac{1}{\mathcal{D}_\mathcal{H}}\sum_{M_k\in\Lambda_n}\delta_{M,M_k},
\ee
where $\Lambda_n=[M_n,M_n+dM]$ and $\mathcal{D}_\mathcal{H}=2j+1$ is the Hilbert space dimension.
$P(M)$ quantifies the probability of finding $M$ in an infinitesimal interval $M\in\Lambda_n$. 

In Fig.~\ref{MDis}(a), we show the histogram of $P(M)$ for the kicked top model 
with $j\in[150,154]$, $\gamma=2.6$, and $\alpha=11\pi/19$. 
Clearly, $P(M)$ behaves as a continuous distribution over the range $M\in[-1,1]$ 
and has two expected sharp clusters around
$M=-1$ and $M=+1$, corresponding to regular and chaotic eigenstates, respectively.
The existence of the intermediate values of $M$ indicates that 
apart from the regular and chaotic eigenstates, there also exist many mixed eigenstates.
To see this, we plot the Husimi function for several eigenstates with different $M$ values in Figs.~\ref{MDis}(c)-\ref{MDis}(d).
Comparing to the classical Poincar\'{e} section in Fig.~\ref{MDis}(b), one can see that the regular eigenstate 
with $M=-1$ is entirely localized in the regular island [Fig.~\ref{MDis}(c)], while the chaotic eigenstate with $M=+1$ exhibits 
a quite uniform distribution over the chaotic sea, as illustrated in Fig.~\ref{MDis}(f).
For the eigenstates with intermediate values of $M$, 
we see the tunneling between different regular island chains [cf.~Figs.~\ref{MDis}(d)], 
as well as between the regular region and chaotic component, as demonstrated in Fig.~\ref{MDis}(e).

Further properties of $P(M)$ are revealed in Fig.~\ref{PMdNk}, where we plot $P(M)$ for several system size ensembles
with different kicking strengths.
Based on these results we make the following observations: 
(i) As the semiclassical limit is approached with increasing $j$, the larger the value of $j$, 
the more the eigenstates move towards the regular or chaotic clusters, regardless of the $\gamma$ value.
(ii) The fluctuations among the intermediate values of $M$ are suppressed as the system size $j$ is increased.
(iii) Increasing $\gamma$ leads to an enhancement in the level of chaos, resulting in $P(M)$ exhibiting a high peak 
around $M=1$ and tiny fluctuations for $-1<M<1$.

 \begin{figure}
  \includegraphics[width=\columnwidth]{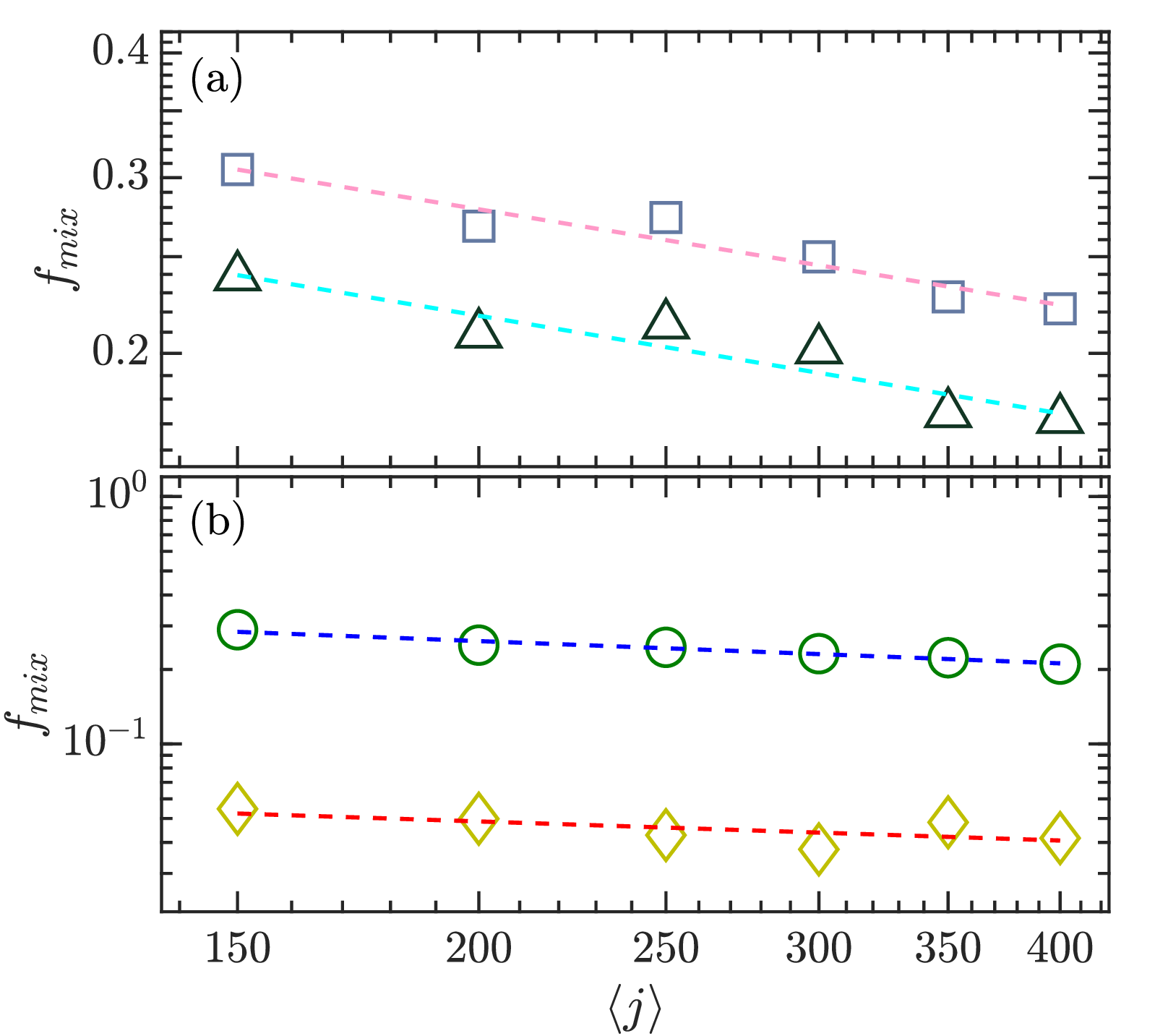}
  \caption{(a): Proportion of mixed states $f_{mix}$ as a function of ensemble averaged system size $\la j\ra$
  for the eigenstates with $M\in[-0.8,0.6]$ (squares) and $M\in[-0.5,0.6]$ (triangles).
  The pink and cyan dot-dashed lines denote the fitting curves of the power law
  $f_{mix}\propto\la j\ra^{-\zeta}$ with $\zeta=0.3184$ and $\zeta=0.3253$, respectively. 
  The kicking strength for this case is $\gamma=2.3$.
  (b): $f_{mix}$ as a function of $\la j\ra$ for $M\in[-0.8,0.7]$ (circles) 
  and $M\in[-0.2,0.2]$ (diamonds) with $\gamma=2.6$. 
  The blue and red dashed lines are the power law fitting curves, $f_{mix}\propto\la j\ra^{-\zeta}$, with
  $\zeta=0.2986$ and $\zeta=0.2561$.
  Other parameter: $\alpha=11\pi/19$. The system size ensemble for each $\la j\ra$ is given by 
  $j\in[\la j\ra-5,\la j\ra+5]$, increasing in step of $1$.}
  \label{MxR}
 \end{figure}

The evolution of $P(M)$ observed in Fig.~\ref{PMdNk} allows us to conclude 
that the relative fraction of the mixed eigenstates decreases with increasing system size. 
To verify this statement and to quantitatively characterize the behaviors observed in Fig.~\ref{PMdNk}, 
we investigate how the proportion of the mixed states varies as a function of system size $j$.
To this end, we choose an interval $\Delta M$ in the range $-1<M<1$, and consider the relative fraction 
of the mixed states belonging to $\Delta M$, defined as
\be
   f_{mix}=\frac{\mathcal{N}^{\Delta M}_{mix}}{\mathcal{D}_\mathcal{H}},
\ee
where $\mathcal{N}^{\Delta M}_{mix}$ is the total number of the mixed states in $\Delta M$ and 
$\mathcal{D}_\mathcal{H}=2j+1$ denotes the Hilbert space dimension.

In Fig.~\ref{MxR}, we plot how $f_{mix}$ evolves as a function of the ensemble averaged system size $\la j\ra$
for different $\gamma$ and $\Delta M$ values.
Overall, we see that irrespective of the values of $\gamma$ and $\Delta M$, 
the decay of $f_{mix}$ with increasing $\la j\ra$ is well described by the power law of the form $f_{mix}\sim\la j\ra^{-\zeta}$.
However, the decay rate, $\zeta$, depends on both $\gamma$ and $\Delta M$.
Besides, when we compare the $y$-axis scale in Fig.~\ref{MxR}(a) with Fig.~\ref{MxR}(b), we observe that 
$f_{mix}$ undergoes a drastic decrease as $\gamma$ is increased, consistent with the results in Fig.~\ref{PMdNk}.
Clearly, with $\gamma$ increasing the chaos increases and thus the number of mixed states decreases as well. 
Furthermore, due to the larger fluctuations in $P(M)$ for small $\Delta M$, the degree of agreement 
between the numerical data and power-law decay decreases with decreasing $\Delta M$. 
But, as the fluctuations in $P(M)$ reduce with increasing system size, 
one can expect a substantial improvement in the quality of the power-law fitting for larger system sizes.  

The power-law decay of the proportion of mixed states has also been observed in billiards,
but the decay exponent is different from our considered model \cite{Lozej2022}. 
A general understanding of the underlying mechanism of such difference 
suggests the need for further exploration.
Nevertheless, the similar decay behavior exhibited by the two different systems leads us to conjecture that
the power-law decay is a universal property of the relative fraction 
of the mixed states, independent of any specific system. 
A theoretical investigation of the decay behavior of the relative fraction of mixed states would be a very interesting topic
for our future work. 

\begin{figure}
  \includegraphics[width=\columnwidth]{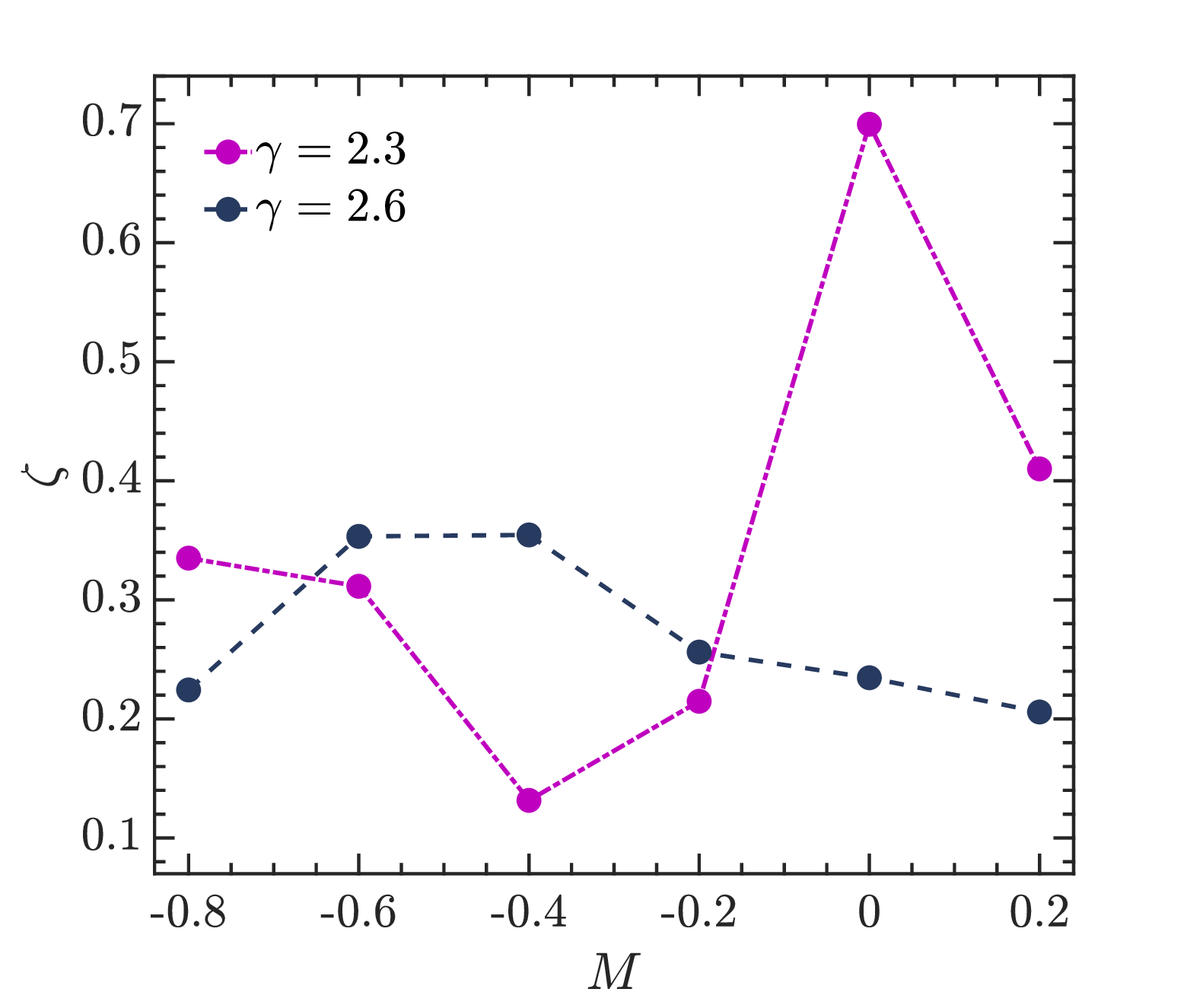}
  \caption{Power law decay exponent $\zeta$ for $M\in[M+\Delta M]$ with $\Delta M=0.4$. 
  The curves correspond to different kicking strengths.
  Other parameter: $\alpha=11\pi/19$.}
  \label{Expfmix}
 \end{figure}

We finally discuss the dependence of the decay exponent $\zeta$ on the different choices 
of intervals with $\Delta M=0.4$.   
The results for two different kicking strengths are plotted in Fig.~\ref{Expfmix}. 
A prominent feature observed in the behavior of $\zeta$ is its larger fluctuations, regardless of the kicking strength.
This stems from the fact that fluctuations in $P(M)$ vary with the choice of intervals, as seen in Fig.~\ref{PMdNk}.
However, the enhancement of the degree of chaoticity with increasing $\gamma$ gives rise to
a remarkable decrease of the fluctuations in the behavior of $\zeta$.
Moreover, as $P(M)$ becomes more smooth as the system size increases, one can expect that
the fluctuations in $\zeta$ will be suppressed for larger system sizes. 
Note that the behavior of $\zeta$ studied here is distinguished from the one revealed in billiard systems, where
$\zeta$ exhibits a rather smooth dependence on the value of $M$ for fixed $\Delta M$ \cite{Lozej2022}.

\section{Conclusions} \label{Fourth}

In summary, we have performed a detailed analysis of the 
properties of the mixed states in the kicked top model.
Being a prototype model for studying quantum chaos, the kicked top model is known for exhibiting a transition to chaos
for both classical and quantum cases \cite{Haake2001} which is unveiled by various chaos indicators 
and shows a good quantum-classical correspondence between them.
For chosen values of kicking strength, the classical phase space of the kicked top model exhibits complex structure with
several regular islands coexisting with a dominating uniform chaotic component. 
This indicates that the eigenstates of the analogous quantum counterpart 
should also have different types of behavior.

To identify the types of the eigenstates, we employ the Husimi function to define the phase space overlap index $M$, 
which measures the degree of overlap of the Husimi function with chaotic and regular regions in the classical phase space. 
The definition of $M$ implies that it will ideally take the value $+1$ for fully chaotic eigenstates, 
while for purely regular eigenstates it equals $-1$. 
However, we have shown that it varies between $-1$ and $+1$ in the near semiclassical limit 
with $-1<M<1$ corresponding to the mixed states
characterized by various tunneling processes among different phase space structures.

Further features of the mixed states are revealed by the probability distribution of $M$.
We have demonstrated that the distribution of $M$ has two peaks at $M=\pm1$, which 
become more sharp with increasing system size (i.~e. approaching the semiclassical limit), 
agreeing with the prediction of the principle of uniform semiclassical 
condensation (PUSC) of Wigner functions (or Husimi functions) \cite{Robnik2019,Robnik2020}.
A quantitative description of the signatures exhibited by the probability distribution of $M$ is provided by
the proportion of the mixed states with associated $M$ belonging to a given interval.  
We have shown that the dependence of the proportion of the mixed states on the system size is well captured
by the power-law decay whose exponent varies for different choices of the interval and kicking strengths. 
Therefore, one can expect that the relative fraction of the mixed states will disappear 
in the far semiclassical limit, confirming the statement of PUSC.  

The same power-law decay of the proportion of mixed states observed in billiards \cite{Lozej2022} 
allows us to conjecture that the proportion of mixed states in other mixed-type systems 
should also decay as a power law in the semiclassical limit. 
A natural extension of the present work is to systematically study the properties of mixed states in quantum
systems with a well-defined classical limit, such as the Dicke model
\cite{QWang2020,Villasenor2021,Emary2003,Magnani2016,Lerma2019,Swan2019,Corps2022}, 
coupled top model \cite{Mondal2020,Mondal2022}, 
and three-site Bose-Hubbard model \cite{Michael2020,Wittmann2022,Nakerst2023}.
Another open topic that deserves investigation is to establish a general theoretical understanding of 
the power-law decay demonstrated by mixed states.
Finally, it is worth pointing out that confirmation of the PUSC in the kicked top model provides further evidence
of the correctness of the Berry-Robnik picture in analysis of spectral statistics of generic quantum systems.

\acknowledgements

This work was supported by the Slovenian Research and Innovation Agency (ARIS) 
under the Grants Nos.~J1-4387 and P1-0306.
Q.~W. acknowledges support from the National Science 
Foundation of China (NSFC) under Grant No.~11805165,
Zhejiang Provincial Nature Science Foundation under Grant No.~LY20A050001.

\bibliographystyle{apsrev4-1}
\bibliography{MixKtop}

\end{document}